\documentclass[conference]{IEEEtran}
\IEEEoverridecommandlockouts
\usepackage{cite}
\usepackage{amsmath,amssymb,amsfonts}
\usepackage{algorithmic}
\usepackage{graphicx}
\usepackage{textcomp}
\usepackage{mathtools}
\usepackage{multirow}
\usepackage[linesnumbered,ruled,vlined]{algorithm2e}
\usepackage{tabularx}
\usepackage{bbm}

\newcolumntype{P}[1]{>{\centering\arraybackslash}p{#1}}
\newcolumntype{C}{>{\centering\arraybackslash}X} 

\newcommand{\beql}[1]{\begin{equation}\label{#1}}
\newcommand{\eeq}{\end{equation}}

\newcommand{\be}{\begin{equation}}
\newcommand{\ee}{\end{equation}}
\newcommand{\ba}{\begin{array}}
\newcommand{\ea}{\end{array}}

\usepackage{color}
\usepackage{xcolor}
\usepackage{soul}

\begin{document}

\title{Adversarial Machine Learning in Wireless Communications using RF Data: A Review
\thanks{This research work is supported by the U.S. Office of the Under Secretary of Defense for Research and  Engineering (OUSD(R\&E)) under agreement number FA8750-15-2-0119 and FA8750-15-2-0120. }
}

\author{\IEEEauthorblockN{Damilola Adesina$^{\dag}$, Chung-Chu Hsieh$^{\ddag}$, Yalin E. Sagduyu$^{\S}$,  and Lijun Qian$^{\dag}$}
\IEEEauthorblockA{$^{\dag}$Center of Excellence in Research and Education for Big Military Data Intelligence (CREDIT Center) \\
 Prairie View A\&M University, Texas A\&M University System, Prairie View, TX 77446, USA\\
 $^{\ddag}$ Center of Excellence in Cyber Security, Norfolk State University,  Norfolk, VA 23504, USA\\
 $^{\S}$ Intelligent Automation, Inc., Rockville, MD 20855, USA \\
Email: dadesina@pvamu.edu, ghsieh@nsu.edu, ysagduyu@i-a-i.com,  liqian@pvamu.edu}
}

\maketitle
\begin{abstract}
Machine learning (ML) provides effective means to learn from spectrum data and solve complex tasks involved in wireless communications. Supported by recent advances in computational resources and algorithmic designs, deep learning (DL) has found success in performing various wireless communication tasks such as signal recognition, spectrum sensing and waveform design. However, ML in general and DL in particular have been found vulnerable to manipulations thus giving rise to a field of study called adversarial machine learning (AML). Although AML has been extensively studied in other data domains such as computer vision and natural language processing, research for AML in the wireless communications domain is still in its early stage.~\textcolor{black}{This paper presents a comprehensive review of the latest research efforts focused on AML in wireless communications while accounting for the unique characteristics of wireless systems. First, the background of AML attacks on deep neural networks is discussed and a taxonomy of AML attack types is provided. Various methods of generating adversarial examples and attack mechanisms are also described. In addition, an holistic survey of existing research on AML attacks for various wireless communication problems as well as the corresponding defense mechanisms in the wireless domain are presented. Finally, as new attacks and defense techniques are developed, recent research trends and the overarching future outlook for AML for next-generation wireless communications are discussed.}
\end{abstract}


\section{Introduction}
\label{sec:introduction}
Machine learning (ML) in general and deep learning (DL) in particular has found rich applications in various domains such as computer vision (CV) and natural language processing (NLP). Motivated by the success in those domains, there has been a growing recent interest in the development of artificial intelligence driven solutions for wireless communications using radio frequency (RF) data \cite{Erpek2020DeepLFWC, Simeone2018very, Wong2020rfml}. For example, convolutional neural network (CNN) models have been used for modulation recognition~\cite{Oshea2016ModRec} and channel decoding~\cite{Liang2018AnIteBPCNN}. In addition to CNN, feed-forward neural network (FNN) and long short term memory (LSTM) models have been used for device fingerprinting and identification~\cite{Jafari2018IoTDF, Jian2020deep, Merchant2018deep, Davasliogludeepwifi}. On the other hand, various deep neural network (DNN) models have been proposed for spectrum sensing and prediction ~\cite{Omotere2018SpecOccPred}, wireless resource management \cite{Sun2017LearnToOpt}, beam prediction for initial access \cite{cousik2020fast} and re-configurable intelligent surfaces \cite{Abuzainab2021}. Although the application of DL-based techniques to wireless communications has shown promising results to solve complex problems for which analytical solutions are either unavailable or computationally infeasible, there are concerns in terms of reliability and robustness of such data-driven methods when compared to traditional signal processing methods in real-world deployments. One of such concerns is the presence of adversaries that may target the training process and/or testing (inference) process of DL.

With the increased adoption of DL, it is inevitable that adversaries will explore launching new attacks particularly on DL models, thus leading to a new paradigm called adversarial machine learning (AML). AML studies learning in the presence of adversaries with the goal of understanding the impact of AML attacks, defending against such attacks, and eventually developing secure DL systems. One type of AML attack (namely adversarial attack or evasion attack) occurs when an adversary crafts small perturbations to the original input of a neural network or develop a means to manipulate the operation of the DL system to cause an error in the inference process. These perturbations are not just white noise samples but they are specifically crafted to produce a vector in the input feature space that is capable of misleading the developed DL model. Classical examples of such attacks include adding small incremental value in the model's gradient direction with respect to the inputs or by solving a constrained optimization problem to produce a vector in the input feature space that is capable of misleading the target DL model. 

\begin{figure}[h]
	 \centering
    	 \includegraphics[width=\columnwidth]{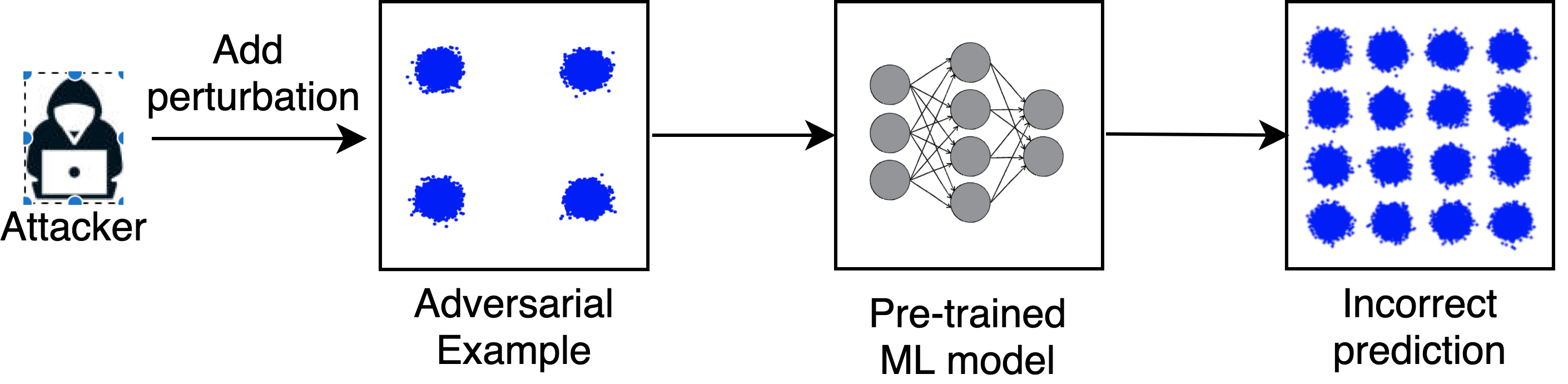}
      	\caption{\textcolor{black}{An example of adversarial attack in wireless communications. An adversary adds perturbation to a test sample to create an adversarial example. Then, the pre-trained ML model wrongly predicts the test sample.}}
    \label{fig:amlexample}
\end{figure}

AML attacks for CV and NLP have been extensively studied~\cite{Szegedy2014IntriguingPO, Vorobeychik2017, Shi2018vulnerability}. A canonical example is the mis-classification of a panda image as gibbon because of the addition of a specifically crafted perturbation to the panda image so as to mislead the ML classifier~\cite{Goodfellow2015ExplainingAH}. Similarly, in the wireless domain, adversarial attacks can lead to mis-classification of signals.~\textcolor{black}{As shown in Figure~\ref{fig:amlexample}, the pre-trained ML model is a classifier for automatic modulation classification. An attacker launches an adversarial attack on the test data in the inference phase by adding a small perturbation to the original QPSK test data sample to form an adversarial example (AdEx). This AdEx causes the mis-classification of a QPSK modulated signal as 16-QAM.}


There have been multiple research efforts that aim to survey AML attacks as well as recent advances to detect and mitigate them in specific areas where DL has found applications such as CV and NLP~\cite{Akhtar2018ThreatofAdAtt,Xu2020AdversarialAA}. This review focuses on approaches to generate, detect and mitigate AML attacks in the wireless domain by accounting for the unique characteristics of wireless communications and their effects on AML attacks.

\subsection{Unique Properties of Wireless Adversarial Attacks}

~\textcolor{black}{Data-driven, intelligent solutions empowered by ML/DL are considered as the key enablers in 5G and 6G wireless communications~\cite{Mahmood6Key2020, Golam2018BDAMLAI, Gustav2020Ericsson, Lin20196GVis}. DL has the strong potential to assist network decisions in achieving optimal resource management and future networks are envisioned to rely on DL-driven agents for functions such as network management automation and optimization of radio interfaces. The evaluation of the extent to which an adversary can affect such a complex system becomes critical in designing future communication systems with DL.  As seen in~\cite{Liao2011SecureML, Barreno2006CanMLSec}, the likelihood of jeopardizing ML/DL models deployed to enhance systems is a challenge that can have dire consequences for ML adoption. For instance, in cyberphysical systems such as autonomous vehicles, adversarial attacks causing incorrect ML/DL decisions may harm lives and properties~\cite{olowo2021ResML, Qayyum2020SecConAVs}. As such, to benefit from the gains of using ML/DL systems, there is an urgent need to ensure security and robustness of ML-based systems in the presence of adversarial attacks. An in-depth study of AML specifically for wireless communication systems is ultimately needed given the unique characteristics and emerging use cases of DL in such systems. Below, we discuss these unique characteristics in the context of AML.}

\begin{itemize}
	\item \textcolor{black}{\textit{Effect of the communication channel:} The wireless communication channel has a major impact on adversarial attacks as they need to be launched over the air to reach the target receivers that host the victim DL models. One aspect is that the channel will impose path loss and phase on adversarial perturbations, and could weaken and/or change the direction of adversarial signals as they travel through wireless channels before reaching the target receivers. For an adversarial attack to succeed over the air, the adversary needs to account for the dynamic nature of channels when crafting the adversarial perturbations \cite{KimChannelAware, kim2020OTAAdAtt, Kim2020AdversarialAW,  Hameed2019TheBD,  restuccia2020generalized}. In addition, the adversary may not have access to the DL model at the target receiver and the process of data gathering by the adversary for training a surrogate model is typically performed through a channel, which implies that the training data used by the adversary is imperfect by default and the effectiveness of the surrogate model trained by the adversary strongly depends on channel effects~\cite{Kim2020surrogatemodel}. Furthermore, it has been shown in \cite{sahay2020frequency} that adversarial attacks crafted to fool DL models trained on time-domain features do not necessarily transfer to DL models trained using frequency-domain features. By leveraging differences in channels, it is also viable to seek the objective of correct signal classification at one receiver while fooling the signal classifier at another receiver  \cite{KimChannelAware, kim2020HowToMake, HameedGlobalSIP, Hameed2019TheBD, berian2020adversarial}.} 
	
	\item \textcolor{black}{\textit{Exploitation and mitigation of channel effects:} The adversarial perturbation may aim to fool a wireless signal classifier at one receiver while the perturbed signal can be still decoded at the intended receiver with minimal loss of reliability \cite{HameedGlobalSIP}. This paradigm can be used to hide wireless signals (ranging from simple modulated signals to complex 5G signals) from eavesdroppers \cite{kim2020HowToMake}. While balancing the objectives of covertness and communication performance, the adversary may aim to preserve not only the bit error rate (BER) as in \cite{Hameed2019TheBD} but also the spectral shape of the perturbed signal as in \cite{Del2020investigating}. In addition, forward error correction (FEC) - a building block in communication systems to detect and correct errors encountered due to noise, channel effects, and hardware impairments - may reduce the effect of adversarial attack while maintaining communication performance with the intended receiver \cite{Del2020effects}. On the other hand, the adversary may use multiple antennas to transmit perturbations to increase the effect of adversarial attacks \cite{Kim2020AdversarialAW}.}
	\item \textcolor{black}{\textit{Indirect influence on training and test data:} A wireless adversary cannot directly manipulate the training or testing data input to a classifier. It also cannot directly query a transmitter's classifier and obtain its classification results as opposed to the typical CV and NLP applications that often rely on API queries. In an over-the-air attack, the wireless adversary needs to monitor the actions in wireless communications and indirectly try to manipulate/influence the outcome of the DL model or initiate actions such as jamming the channel during sensing and data transmissions~\cite{Sagduyu2019AdDLOTA}.}

 	\item \textcolor{black}{\textit{Heterogeneity in feature representation:} The coexistence of various communication systems introduces heterogeneity as a more diverse and complex feature representation in radio data~\cite{Lin2020ThreatAdAtt}. This significantly affects the effectiveness of crafted perturbations. In this context, the adversary may also aim to fool the wireless signal classifier at multiple receivers (each possibly belonging to a different network) by transmitting a signal perturbation and relying on the broadcast nature of omnidirectional transmissions to reach and jointly affect multiple classifiers. Then, the perturbation needs to be determined by accounting for multiple channels to different receivers \cite{KimChannelAware}. In addition to dynamic channels, it is also possible that network traffic is of stochastic nature or traffic needs to be adapted to dynamic channel effects. Then, the generation of perturbations needs to be jointly considered with queue stability with the extended goal for the adversary to reduce the stable throughput (the maximum achievable throughput while keeping the packet queues stable \cite{queuestability}).} 
\end{itemize}
\textcolor{black}{These unique properties necessitate the review of the current attacks and mitigation methods of AML specifically for wireless communications as the methods presented in other domains such as CV and NLP may not fit to the characteristics of wireless medium. For instance, the adversary can directly query an API for CV applications without the need to send perturbations over a channel.}

The design of DL algorithms for wireless communications must consider the new security issues such as identification and mitigation of adversarial signals \cite{sagduyu2020whenwire}. Due to dynamic properties of RF data, it is a non-trivial task to identify adversarial examples in RF data. DL methods rely on the premise that the training data represents the distribution of the underlying data generation process such that the trained model generalizes well on test data. However, this premise does not necessarily apply in the adversarial scenarios as these distributions change significantly due to adversarial effects that are not straightforward to anticipate or predict~\cite{Roy2019MLinAdRFEnviron,Adesina2020RobustDRF,Adesina2019PracticalRF, shi2019deepDyspan}. With such adversarial attacks, wireless communications systems such as 5G and 6G with design components based on DL will be susceptible to disruptions, performance losses, and eventually failures. Therefore, it is critical to analyze the emerging wireless attack surface due to AML, and reduce and ultimately eliminate the impact of adversaries employing the AML attack techniques.

The rest of this paper is organized as follows. In Section~\ref{sec:relatedwork}, previous works on attacks on wireless communication systems as well as AML on other domains are presented. Section~\ref{sec:DLapproach} presents the background on AML detailing the development of DNNs and adversarial perturbations. Section~\ref{sec:adMLattacks} explores different types of AML attacks and provides a taxonomy of AML attacks in the wireless domain. Section~\ref{sec:adversarialexamplesinWireComm} discusses the generation of AML attacks. The review of the work on AML grouped under various wireless communication problem in done in Section~\ref{sec:amlreview}. In addition, AML attack detection and mitigation methods are presented in Section~\ref{sec:detectmiti} while the future outlook of AML in wireless communications is presented in Section~\ref{sec:futureoutlook} and Section~\ref{sec:conclusions} the paper. The symbols and notations as well as the definition of some important terms used in this review are summarized in Section~\ref{sec:symbNot} and Section~\ref{sec:termsndef} respectively.

\subsection{Symbols and Notations}
\label{sec:symbNot}
\begin{minipage}{0.1\textwidth}
\begin{tabular}{>{\raggedright}p{1cm} }
 $X$\\
 $x$ \\
 $x^*$\\ 
 $Y$\\  
 $y$\\
 $y^\prime$\\ 
 $r_x$\\

\end{tabular}
\end{minipage}
\begin{minipage}{0.1\textwidth}
\begin{tabular}{>{\raggedright}p{5cm} }
total data points \\
a data point  \\
adversarial example \\ 
class (label) for dataset \\  
class for a data point  \\
particular target label\\ 
adversarial perturbation \\
\end{tabular}
\end{minipage}

\begin{minipage}{0.1\textwidth}
\begin{tabular}{>{\raggedright}p{1cm} }
$w$\\ 
$J$\\ 
$\alpha$\\ 
$\epsilon$\\
$C$\\ 
$\bigtriangledown$\\ 
$f_w$\\

\end{tabular}
\end{minipage}
\begin{minipage}{0.1\textwidth}
\begin{tabular}{>{\raggedright}p{5cm} }
model parameters\\ 
cost function \\ 
step size \\ 
small positive number\\
number of classes\\ 
gradient operator \\ 
neural network classifier function \\
\end{tabular}
\end{minipage}

\subsection{Terms and Definitions}
\label{sec:termsndef}

\begin{table}[htbp] \scriptsize
\centering
\begin{tabular}{m{2.5cm} m{5cm}}
\hline

\textbf{Terms} & \textbf{Definitions} \\ \hline

\textcolor{black}{\textbf{Adversary}} & \textcolor{black}{An agent that adds perturbation to clean data with the aim of fooling ML models.}\\ \hline

\textcolor{black}{\textbf{Adversarial Perturbation}}& \textcolor{black}{The perturbation (noise or disturbance) signal generated by an adversary and added to clean data to alter it so as to fool an ML based system such as a classifier.} \\ \hline 

\textcolor{black}{\textbf{Adversarial Example}} & \textcolor{black}{A sample of the data that has been altered by adding a perturbation to fool an ML based system.} \\ \hline

\textcolor{black}{\textbf{Adversarial Attack}} & \textcolor{black}{The process by which an adversary injects the adversarial example to undermine the ML-based system.} \\ \hline

\textcolor{black}{\textbf{Adversarial Machine Learning}} & \textcolor{black}{Learning in the presence of adversaries with the goal of understanding the impact of AML attacks, defending against such attacks, and eventually developing secure DL systems.} \\ \hline

\textcolor{black}{\textbf{Adversarial Training}} & \textcolor{black}{A defense mechanism where perturbed samples are added to the training data to boost model robustness against adversarial attacks.} \\ \hline

\textcolor{black}{\textbf{Transferability}} & \textcolor{black}{The ability of an adversarial example to retain its potency when transferred between different models.} \\ \hline

\textcolor{black}{\textbf{Robustness}} & \textcolor{black}{The ability of a model to remain reliable under changing conditions including adversarial attacks.} \\ \hline

\end{tabular}
\label{table:def}
\end{table}

\section{\textcolor{black}{Related work}}
\label{sec:relatedwork}
\textcolor{black}{DL has been extensively used in wireless communications as outlined in numerous survey papers \cite{Erpek2020DeepLFWC, Simeone2018very, Wong2020rfml, Sun2019AppOfMLinWcomm, Dai2020DL4WireComm, kulin2020survey, Klaine2017ASurMLTech, Nguyen2021EnaAI, JAGANNATH2019101913, Chen2019ANNMLwirecomm, Jiang2016MLPara, Li2021MLinWireComm}. These papers highlight various models and applications of DL in wireless communication systems. Specifically for AML in wireless communications, Table~\ref{table:modelsinML} shows the types of DL models used in various related works.}

\begin{table}[htbp] \scriptsize
\caption{DL models for Adversarial Machine Learning in Wireless Communications.}
\centering
\begin{tabular}{|m{3.1cm}|m{4.9cm}|m{2cm}|m{1.5cm}|m{2.5cm}|m{5cm}|}

\hline

\textbf{DL Models} & \textbf{Papers} \\ 
\hline\hline
Feedforward Neural Network & ~\cite{Sagduyu2019IoTNetSec, kim2020OTAAdAtt, Shi2018AdDL4CogRaSec, Shi2018SpecDaPoiAML, Sadeghi2019PhyAdAttAuto, Shi2020OTAMIA, shi2021membership}\\ \hline

Convolutional Neural Network & ~\cite{Lin2020ThreatAdAtt, Sadeghi2019AdAttDLRSC, Kim2020AdversarialAW, Del2020investigating, Kim2020surrogatemodel, Del2020effects, filipovic2019MitAdEx, Sadeghi2019PhyAdAttAuto, Bair2019OnLimofTar,Ke2019AppAdExComm, kim2020HowToMake, Hameed2019TheBD, Hameed2019CommunicationWI, kokaljfilipovic2019adversarial, Davaslioglu2019TroAttWSC, Liu2020AdvAttonDL, KimChannelAware, restuccia2020generalized, Flowers2020EvaAdEx,  restuccia2020hacking, fili2019TargetAdEx, Sagduyu2019AdDLOTA} \\ \hline

Recurrent Neural Network & ~\cite{karunaratne2020penetrating} \\ \hline

Reinforcement Learning & ~\cite{Zhong2020adversarial, Wang2020adversarial, Shi2021HowTA, Shi2021Adversarialaa, Wang2020defense} \\ \hline

\end{tabular}
\label{table:modelsinML}
\end{table} 

~\textcolor{black}{There are various surveys in related areas, for instance,~\cite{Deog2017SecAttIoT, shahzad2017survey, Jaitly2017SecVul, Vadlamani2016JamAtt, Wu2018SurPhyLaySec, Stellios2018SurIotatt, Zou2016ASurWireSec, ARIF2019100179, Dewal2018SecAttWSN, Manaswi2019SurAttMitiSDN} surveyed attacks on wireless communications systems,~\cite{Sagduyu2019IoTNetSec, kim2020OTAAdAtt, Shi2018AdDL4CogRaSec, Shi2018SpecDaPoiAML, Sadeghi2019PhyAdAttAuto, Flowers2020EvaAdEx,  restuccia2020hacking, fili2019TargetAdEx, Del2020investigating, Kim2020surrogatemodel, Del2020effects, Zhong2020adversarial, Wang2020adversarial, Shi2021HowTA, Shi2021Adversarialaa} used various ML models for wireless communications while adversarial ML in other domains such as computer vision and NLP was discussed in~\cite{bhambri2020survey, Qayyum2020SecConAVs, Yuan2019AdExAttDefDL, Ivana2017SurPotSec, Barreno2006CanMLSec, Biggio2018WildPatterns, Akhtar2018ThreatofAdAtt, Huang2011AdvML, chakraborty2018adversarial, NOWROOZI2021102092, machado2020adversarial, Serban2020AdExOnObjRec, sun2020adversarial}. To the best of our knowledge, there is no thorough survey available for AML in wireless communication systems.}

Traditional attacks on communication systems such as primary user emulation, jamming, eavesdropping attacks, and spectrum sensing data falsification have been extensively studied in~\cite{Zou2015SecPhyLay,Zou2016SurWireSec, sagduyu2014securing, sagduyu2011jamming, sagduyu2010wireless}. To further secure communication systems, ML/DL techniques themselves can be used to detect and/or mitigate such traditional attacks~\cite{Rajendran2018saife, Van2019jam, Abuzainab2019qos}. On the other hand, AML has been reviewed and surveyed for various applications such as CV~\cite{Akhtar2018ThreatofAdAtt}, autonomous vehicles~\cite{Qayyum2020SecConAVs}, and cyber security~\cite{Zhou2019survey}. \textcolor{black}{\cite{WANG201912, li2018security, chakraborty2018adversarial, OZDAG2018152, Liu2018ASurSecThrDef} discussed the foundations of AML and provided detailed discussion on attack and defense strategies as well as relevant perspectives for designing more secure ML models.~\cite{PITROPAKIS2019100199} focused on the taxonomy of attacks to identify areas of action for researchers while~\cite{silva2020opportunities} focused on adversarial example generation and defense mechanisms.~\cite{sun2020adversarial} provided a unified formulation for AML on graph data and compared attacks methods and defense strategies. Furthermore,~\cite{bhambri2020survey, NOWROOZI2021102092, Serban2020AdExOnObjRec, machado2020adversarial, Qayyum2020SecConAVs, wiyatno2019adversarial} provided various perspectives on AML in the CV domain. For instance~\cite{Serban2020AdExOnObjRec} focused on AML in object recognition,~\cite{machado2020adversarial} focused on AML in image classification and~\cite{Qayyum2020SecConAVs} focused on AML in autonomous Vehicles. These works presented a taxonomy of attacks, defense strategies and an outlook to the future.} The scope of this review differs from previous works as it explores AML attacks that directly target the growing ML/DL applications in wireless communications.  \textcolor{black}{To the best of our knowledge, this is the first review work that focuses on AML in wireless communications by discussing the unique properties of wireless communications and its effects on AML.} These AML attacks are effective because they operate with low spectrum footprint, and therefore they are stealthy (hard to detect) and energy-efficient~\textcolor{black}{~\cite{olowo2021ResML, sagduyu2020whenwire, Shi2021HowTA, Liu2020AdvAttonDL, restuccia2020hacking}}.~\textcolor{black}{For instance, \cite{Shi2018AdDL4CogRaSec} showed that the DL-based exploratory attack is more effective in reducing the system throughput and more energy efficient than traditional random jamming attacks. The random jamming attack jams the transmission channel in some randomly selected instances while the DL-based exploratory attack jams in carefully selected time slots, thus using less energy.} Due to the unique properties of wireless communications, it is essential to understand and quantify the impact of AML attacks on wireless communications and develop defense mechanisms that are tailored for the wireless domain.

\section{Background of Deep Adversarial Learning}
\label{sec:DLapproach}

\subsection{Deep Neural Network}
Supervised learning using a DNN is expressed as a mapping $f$ that models the mapping from the input data to a class (label). To learn this input-output relationship, a weight matrix $w$ is estimated. A loss function $l(x, y, w)$ is computed as a point-wise measure of error between the model prediction $\hat{y}= f(x)$ and the observed ground truth $y$, where $x \in X$ is a sample data point in data $X$ and $y \in \mathcal{C}$ is a class in $\mathcal{C}$. To estimate $w$, a cost function $J(w)$, which is the average loss over all~\textcolor{black}{$(n)$} training data samples in $X$, is computed as~\cite{Goodfellow-et-al-2016}

\begin{equation}
\label{eq:DL3}
 \textcolor{black}{J(w) \equiv  J(X, y, w) = \frac{1}{n}\sum \limits_{(x_i, y_i)} l(x_i, y_i, w).}
\end{equation}


The model is derived by minimizing the cost function $J(w)$ as 
\begin{equation}
\label{eq:DL4}
\operatorname*{argmin}_{w \epsilon \mathbb{R}^n} J(w).
\end{equation}
Then, the DNN classifier indexed by the DNN weights $w$ is determined as $f_w:X \rightarrow \mathbb{R}^C$, where $C$ is the number of classes in $\mathcal{C}$.

\begin{table*} [ht] \scriptsize
\caption{Summary of previous work on AML attacks in wireless communications.}
\centering
\begin{tabular}{|l|l|l|l|c|c|c|c|}
\hline
\multicolumn{1}{|c|}{\textbf{Categorization}}& \multicolumn{1}{|c|}{\textbf{Attack Type}}& \multicolumn{1}{|c|}{\textbf{Description}} & \multicolumn{1}{|c|}{\textbf{Paper}} \\
\hline
\multirow{3}{*}{Attack Type}&  Exploratory&  Fathom the inner workings of model  &~\cite{Shi2018AdDL4CogRaSec, Erpek2019DLforLM, Sagduyu2019IoTNetSec, Shi2018SpecDaPoiAML, Shi2020OTAMIA, Luo2020attackers, ZluoPartialAttack, Sagduyu2019AdDLOTA, Zhong2020adversarial, Kim2020surrogatemodel, Shi2021HowTA, Shi2021Adversarialaa} \\ \cline{2-4}

& Evasion & Fool ML models &~\cite{fili2019TargetAdEx, Bair2019OnLimofTar, Sadeghi2019AdAttDLRSC, Flowers2020EvaAdEx, Hameed2019CommunicationWI, Lin2020ThreatAdAtt, kim2020OTAAdAtt, Shi2018SpecDaPoiAML, Sadeghi2019PhyAdAttAuto, Sagduyu2019IoTNetSec, Sagduyu2019AdDLOTA, Kim2020AdversarialAW, fili2019AdExinRF,  kim2020HowToMake,  Ke2019AppAdExComm, KimChannelAware, Del2020effects, Del2020investigating, Hameed2019CommunicationWI, karunaratne2020penetrating} \\ \cline{2-4} 

& Causative & Manipulate the training process &~\cite{Shi2018SpecDaPoiAML, Sagduyu2019AdDLOTA, Sagduyu2019IoTNetSec}   \\ \hline

\multirow{2}{*}{Adversary's Goal}  & Targeted & Cause specific errors & ~\cite{fili2019TargetAdEx, Bair2019OnLimofTar, kim2020OTAAdAtt, Kim2020AdversarialAW, kim2020HowToMake, KimChannelAware, Kim2020surrogatemodel}  \\ \cline{2-4}
& Non-Targeted & Cause any errors &~\cite{kim2020OTAAdAtt, restuccia2020generalized, KimChannelAware, Liu2020AdvAttonDL, Bair2019OnLimofTar, karunaratne2020penetrating}  \\ \hline

\multirow{3}{*}{Amount of Knowledge} & White-box & Full knowledge & ~\cite{Sadeghi2019AdAttDLRSC, kim2020OTAAdAtt, Liu2020AdvAttonDL, Sadeghi2019PhyAdAttAuto, Kim2020AdversarialAW, kim2020HowToMake, restuccia2020generalized, KimChannelAware, Kim2020surrogatemodel, karunaratne2020penetrating, Bair2019OnLimofTar} \\ \cline{2-4}
& Black-box & No knowledge &~\cite{Lin2020ThreatAdAtt, Sadeghi2019AdAttDLRSC, kim2020OTAAdAtt, restuccia2020generalized, Sadeghi2019PhyAdAttAuto, Luo2020attackers, fili2019TargetAdEx,KimChannelAware, Wang2020adversarial, Wang2020defense, Shi2020OTAMIA, shi2021membership}  \\ \cline{2-4}
& Gray-box & Limited knowledge &~\cite{filipovic2019MitAdEx, Wang2020adversarial} \\ \hline
\end{tabular}
\label{table:modelum}
\end{table*}

\subsection{Adversarial Attack on Deep Neural Network}
\label{sec:AdattackDNN}
As an example of AML attacks, we describe next how the evasion or the so-called adversarial attack works. We will discuss other types of AML attacks in Section~\ref{sec:adMLattacks}. For every $x \in X$, $f_w(x)$ assigns a  label of the form $ \hat{y} = \operatorname*{argmax}_{y \in \mathcal{C}} f_w (x, y)$, where $f_w(x, y)$ corresponds to the output of $f_w$ for class $y$ in $\mathcal{C}$. An adversarial attack on the  classifier $f_w$ aims to modify the classified input $x$ with perturbation $r_x$ such that $f_w(x+r_x) \neq f_w(x)$. For the stealth nature of this attack, it is required that the perturbation $r_x$ should be small such that the difference of $x+r_x$ from the unperturbed input data $x$ remains small. Typically, this requirement is relaxed to constraint $|| r_x ||_p \leq \epsilon$, where $\epsilon$ is a small positive number and $|| \cdot||_p$ represents the $l_p$ norm with $p \in \{1, 2, \infty \}$ denoting the type of norm. The adversarial perturbation $r_x$ for $x$ and $f_w$ is determined by solving the following optimization problem:
\begin{eqnarray}
\label{eq:D2bb}
 &\min\limits_{r_x} &|| r_x ||_p  \\ \nonumber
&s.t.& f_w(x) \neq f_w(x + r_x).
\end{eqnarray}

In the context of wireless communications, the $l_2$-norm is a natural choice for $r_x$ as it accounts for the perturbation (signal) power~\cite{Sadeghi2019AdAttDLRSC}~\cite{Liu2020AdvAttonDL} compared to other norms such as the $l_1$-norm that has been used in the CV domain to deter the variation from human perception~\cite{Liu2020AdvAttonDL}.

\section{Categorization of Adversarial Machine Learning Attacks on Wireless Communications}
\label{sec:adMLattacks}

\begin{figure}[h]
	 \centering
    	 \includegraphics[width=\linewidth]{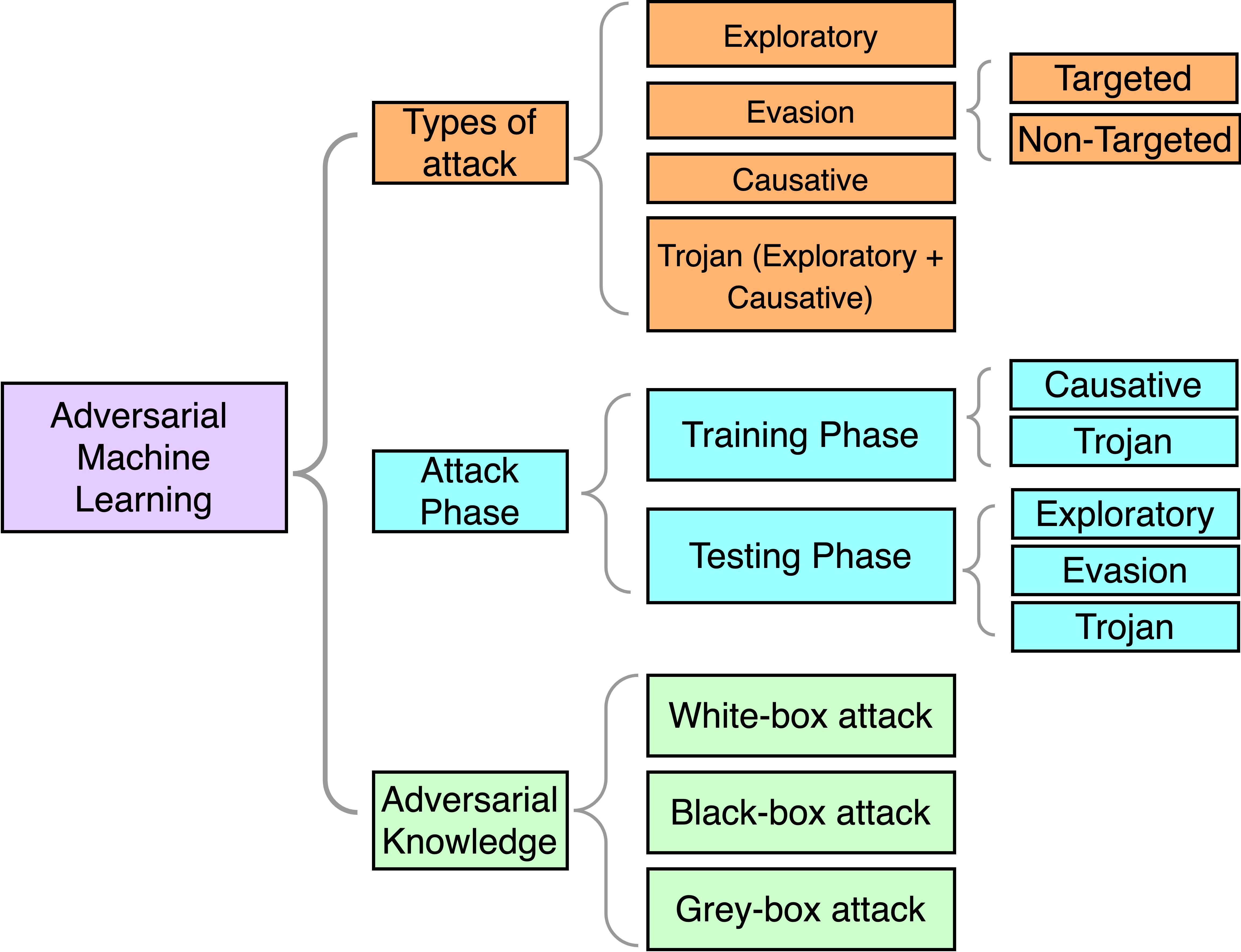}
      	\caption{\textcolor{black}{Categorization of adversarial machine learning.}}
    \label{fig:amlcategory}
\end{figure}

 \textcolor{black}{We present a categorization of AML attacks to aid the understanding of key concepts of AML and provide an overview of AML attacks from various perspectives borrowed from~\cite{Sagduyu2019IoTNetSec, Shi2018SpecDaPoiAML, kim2020OTAAdAtt, Flowers2020EvaAdEx, Shi2018SpecDataPoi, Shi2018AdDL4CogRaSec, Kwon2018FoolNNMin, Lin2020ThreatAdAtt} for wireless communications and~\cite{Yuan2019AdExAttDefDL, Akhtar2018ThreatOfAdvatt, Biggio2018WildPatterns, chakraborty2018adversarial, Papernot2018SoK, Huang2011AdvML, Barreno2006CanMLSec, Liu2018ASurSecThrDef} for other domains especially CV. In particular, we discuss adversarial attacks in wireless communications under three main categories (see Fig.~\ref{fig:amlcategory}) that are defined according to the influence - the type of attack, the attack phase, and the amount of knowledge the adversary has about the victim model. Table \ref{table:modelum} lists previous works of AML attacks on wireless communications under these categories.}


\subsection{Categorization Based on Influence - Types of Attack}
\label{sec:adMLattacksinfluence}

\begin{itemize}
	\item \textit{Exploratory attacks}:  Also called inference attacks, exploratory attacks seek to fathom the inner workings of ML algorithms/models by collecting training data and imitating the ML model functionally with similar types of inputs and outputs, namely building a surrogate (shadow) model~\cite{Tramer2016stealing, Shi2017HowToSteal}. The exploratory attack is typically the leading step prior to subsequent attacks as it aims to ``explore" the victim model using techniques such as active learning~\cite{Shi2018active} or augment the limited information with generative adversarial networks (GANs)~\cite{ShiGAN2018}. Adversarial attacks designed with the surrogate model are known to transfer to the target model (known as the transferability property \cite{Papernot2017PracBBAtt}). In a wireless communications scenario, the adversary can learn the transmission patterns of the victim communication system by observing the spectrum over the air. The effect of wireless channels on surrogate models was studied in \cite{Kim2020surrogatemodel}.

	\item \textit{Evasion attack}: Also called adversarial attacks, evasion attacks aim to fool ML models into making wrong decisions by manipulating the input test data \cite{Szegedy2014IntriguingPO, Goodfellow2015ExplainingAH}. Evasion attacks have been applied to wireless communications in terms of fooling classifiers used for spectrum sensing \cite{Sagduyu2019AdDLOTA}, modulation recognition \cite{Sadeghi2019AdAttDLRSC}, autoencoder-based end-to-end communication systems \cite{Sadeghi2019PhyAdAttAuto}, and channel state information (CSI) feedback for massive MIMO \cite{Liu2020AdvAttonDL, manoj2021AdvAttDLBasPowAll} and initial access in directional communications \cite{kim2021adversarial}. One challenge in the wireless domain is that the adversarial perturbations observed by the target classifier are received subject to channel effects such that these perturbations need to be crafted by accounting for channel effects \cite{KimChannelAware, Hameed2019TheBD, restuccia2020generalized, restuccia2020hacking}. These attacks are typically stealthier and more energy-efficient than traditional jamming attacks (such as the one that solely aim to cause interference to data transmission, e.g., \cite{xu2005feasibility, sagduyu2011jamming,sagduyu2010wireless}) as they only need to transmit low-power signals over a short period of time to confuse the ML algorithms in their decision making.

	\item \textit{Causative attack}: Also called poisoning attacks, causative attacks aim to manipulate the training process of ML models by injecting vulnerabilities such as false training data to the ML models \cite{Biggio2012poisoning, Shi2017evasion}. Causative attacks are effective against DNNs as they are highly susceptible to inaccuracies in training data. An example of a causative attack in wireless communications is spectrum poisoning where the adversary exploits the (re)training process of the ML classifier so that the ML model is poorly (re)trained. Examples of causative attacks in wireless communications include attacks on spectrum sensing \cite{Sagduyu2019AdDLOTA}, cooperative spectrum sensing \cite{Luo2020attackers}, and IoT systems \cite{ZluoPartialAttack, LowCostAttack}. 

\begin{figure*}[htbp]
	 \centering
    	 \includegraphics[width=14cm, height=8cm]{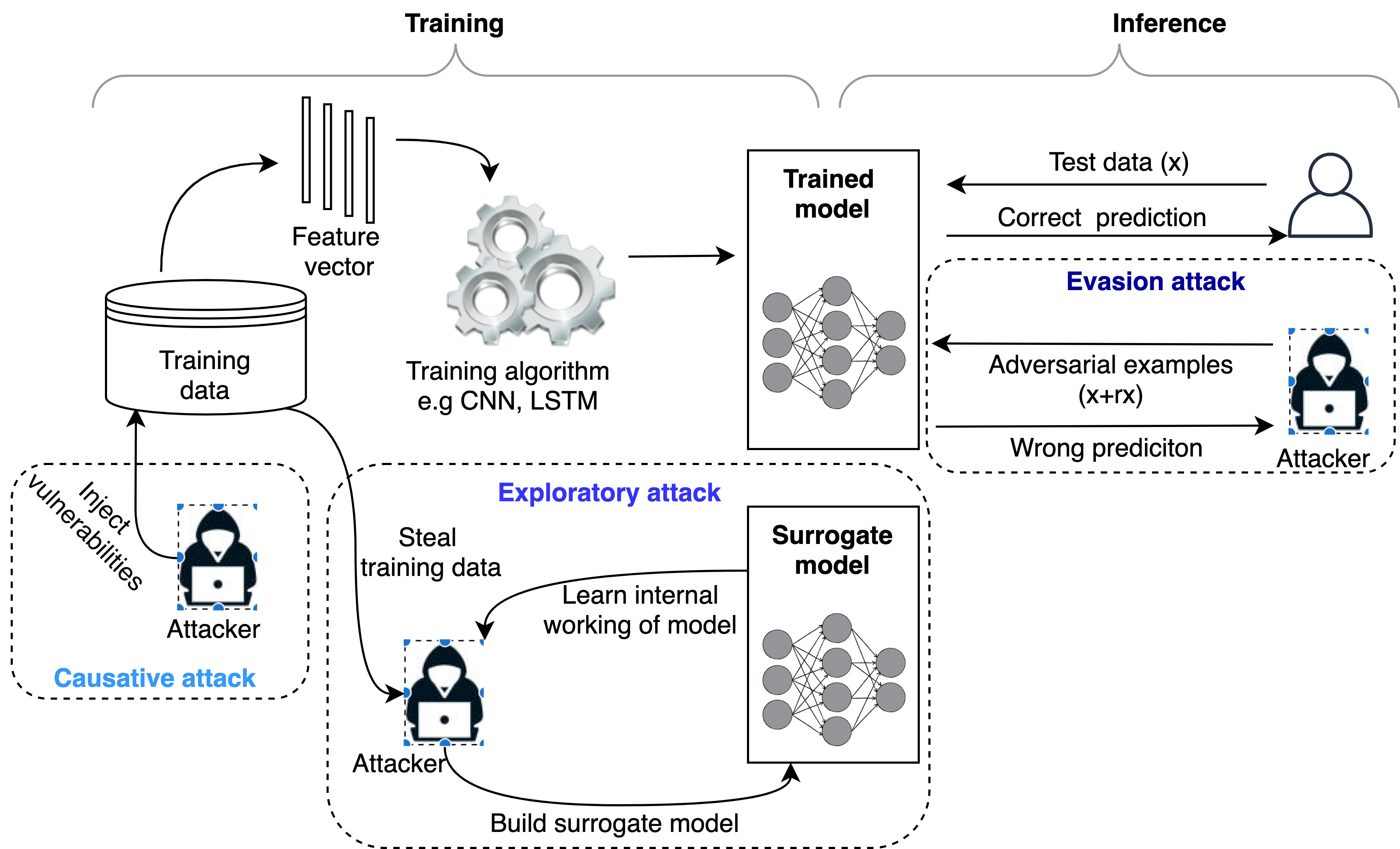}
      	\caption{\textcolor{black}{Attacks in adversarial machine learning~\cite{He2020ToSecTh}}.}
    \label{fig:attacktypes}
\end{figure*}

	\item \textit{Trojan attacks}: Also called backdoor attacks are combinations of evasion and causative attacks, where the adversary injects triggers (backdoors) to training data and then activates them for some input samples in test time \cite{Wang2019neural,Chen2018detecting}. Trojan attack has been considered against wireless signal classifiers in \cite{Davaslioglu2019TroAttWSC} by formulating controlled phase shifts as backdoors and adding them into the transmitted and received RF data samples. 
\end{itemize}

\textcolor{black}{Figure~\ref{fig:attacktypes} gives a pictorial representation of AML attack models and shows how the different attacks work, as well as the general procedure for attack formation.}

\subsection{\textcolor{black}{Categorization Based on Attack Phase.}}
\begin{itemize}
	\item{\textit{\textcolor{black}{Training phase:}}}~\textcolor{black}{The causative and trojan attacks occur in the training phase of the model development. See Section~\ref{sec:adMLattacksinfluence} for discussion on these attacks.}
	\item{\textit{\textcolor{black}{Testing phase:}}}~\textcolor{black}{The exploratory attack, evasion attack and trojan attack occur in the test time. See Section~\ref{sec:adMLattacksinfluence} for discussion on these attacks.}
\end{itemize}

\subsection{Categorization Based on The Amount of Knowledge The Adversary Has About The Victim Model}
\begin{itemize}
	\item \textit{White-box attack}: The adversary knows the training data, architecture, algorithm and/or optimization techniques such that it has full access to the trained model $f$ and knows the input at the classifier. For an over-the-air white-box attack in  wireless communications, the adversary also needs to know the channel between the adversary and the receiver, since the channel affects the perturbation perceived at the receiver~\cite{KimChannelAware, Hameed2019TheBD, restuccia2020generalized, Kim2020surrogatemodel}. 

	\item \textit{Black-box attack}: The black-box attack is a more realistic and more relaxed model for many security threats~\cite{Kurakin2017AdversarialEI} as the adversary has neither knowledge about nor access to the training data and/or the trained model $f$. In that case, the adversary tries to deduce information from the returned results of the model. Black-box attacks typically use a surrogate (shadow) model, which is trained to perform an identical task as the target network during the inference attack. In wireless communications, the surrogate model strongly depends on the channel characteristics with respect to the adversary and the victim system \cite{Kim2020surrogatemodel}. 

	\item \textit{Gray-box attack}: In a gray-box attack, the adversary has some knowledge of the data/algorithm as well as limited access to the model $f$. One example from the wireless domain is not knowing the exact channel but knowing an estimate, the distribution or some statistics of the channel, e.g.,~\textcolor{black}{Rayleigh} vs. Rician fading, the value of pathloss exponent or the value of signal-to-noise-ratio (SNR), etc.~\cite{KimChannelAware}.  
\end{itemize}

\subsection{Categorization Based on Type of Adversary's Goals}
\begin{itemize}
	\item \textit{Targeted attack}: The adversary tries to produce the perturbation $r_x$ that will cause specific errors in the output of a receiver by targeting a signal $x$ to generate errors such that $f_w(x+r_x) = y'$ for a particular target class $y'$. For example, consider a 4-class modulation classification problem with BPSK, QPSK, QAM 16 and QAM 64 as classes and consider the adversary that seeks to fool the modulation classifier into classifying a signal modulated with QAM 16 as a QAM 64.
	\item \textit{Non-targeted attack}: The adversary tries to produce the perturbation $r_x$ that will cause errors in the output of the algorithm independent of the classes thereby reducing the confidence in the algorithm due to decreased accuracy. In our previous example of a 4-class modulation classification problem with BPSK, QPSK, QAM 16 and QAM 64 as classes, the adversary seeks to fool the modulation classifier into classifying a signal modulated with QAM 16 as any other available class such as BPSK, QPSK, or QAM 64, i.e., $f_w(x+r_x) \neq y$ for any $y = f(x)$.
\end{itemize}

\subsection{Other Types of Categorization}
Some other types of categorization for AML attacks include categorization based on the frequency of attack and categorization based on the degree of freedom the adversary has in accessing the system's input.
\begin{itemize}
	\item In the attack frequency category~\cite{Lin2020ThreatAdAtt}, there is a one-step attack that computes the gradient of the loss function at a time to create perturbations. There is also an iterative attack that performs the same computation as a one-step attack multiple times to create the perturbations. This is enabled by perpetual access to the model and can be computationally intensive and costly.

	\item Based on the adversary's degree of freedom to the input of the system, there may be physical and digital attacks. In a physical attack, there is an indirect application of the input to the model which is different from the digital attack where the adversary can explicitly design the input of the model~\cite{Sadeghi2019PhyAdAttAuto}~\cite{Kurakin2017AdversarialEI}. We can also remove the requirement for the adversary to know the input test samples and craft input-agnostic adversarial attacks, also called Universal Adversarial Perturbation (UAP) \cite{Moosavi2017universal}. UAP attacks can be built against wireless signal classifiers, where the adversary does not need to know about the transmitted signal \cite{Sadeghi2019AdAttDLRSC, KimChannelAware}. It is also possible to limit attacks to certain users, e.g., evasion attacks to fool classifiers only at a subset of receivers \cite{KimChannelAware} and causative attacks launched from a selected subset of IoT devices to manipulate the training process at a fusion center \cite{ZluoPartialAttack}.
\end{itemize}

\textcolor{black}{These attack types may overlap in the description of an attack. For example, the authors in~\cite{kim2020OTAAdAtt} studied both targeted, white-box evasion attack and non-targeted white-box evasion attacks for DL-based modulation classification.}


\section{\textcolor{black}{Adversarial Machine Learning Attacks in Wireless Communications}}
\label{sec:adversarialexamplesinWireComm}

In this section, we describe the approaches to generate attacks in AML, including evasion attacks, exploratory attacks and causative attacks.

\subsection{Generation of Evasion Attacks}

Evasion attacks to generate adversarial examples are characterized as subtly crafted imperceptible perturbations $r_x$ added to the original inputs $x$ of the ML model such that the input to the classifier is $x' = x + r_x$ by solving the optimization problem (\ref{eq:D2bb}) for $r_x$. In the wireless domain, the problem is changed by setting with $x' = h_{t,r} x + h_{t,a} r_{x} + n$, where $x'$ is the received signal, $x$ is the transmitted signal, $r_{x}$ is the transmitted perturbation, $h_{t,r}$ is the channel gain from the transmitter to the receiver, $h_{t,a}$ is the channel gain from the transmitter to the adversary, and $n$ is the receiver noise. 

It is difficult to solve (\ref{eq:D2bb}) due to the non-linearity of the DNN mapping. Instead, we can approximate the solution. This is achieved by maximizing the loss function $L(w, x, y)$ used in the training process of the classifier $f_w$ (e.g., the cross-entropy loss function is $L(w, x, y) = -\log(1 + \exp({-f_w(x,y)}))$ subject to the condition that the perturbation is upper bounded. Thus, the adversarial perturbation $r_x$ can be found by solving
\begin{eqnarray}
\label{eq:lossapprox}
&\max& L(w, x+r_x, y) \\ \nonumber
&s.t.& \min || r_x ||_p \leq \epsilon,
\end{eqnarray}
where $\epsilon$ is the upper bound on the adversarial perturbation.
Various methods have been proposed to approximately solve ~(\ref{eq:lossapprox}), thus generating adversarial examples. These methods include but are not limited to one-step gradient-based method such as Fast Gradient Sign Method (FGSM) \cite{Goodfellow2015ExplainingAH}, and its iterative variants such as Basic Iterative Method (BIM) \cite{Kurakin2017AdversarialEI}, Projected Gradient Descent (PGD) \cite{Madry2018TowardsDL}, Momentum Iterative Method \cite{Dong2018BoostAdAttM}, Box Constrained Limited-Memory Broyden–Fletcher–Goldfarb–Shanno (L-BFGS) \cite{Szegedy2014IntriguingPO}, and Carlini \& Wagner (C\&W). We describe some of these methods that have been used to generate adversarial attacks on wireless communications.

\subsubsection{Fast Gradient Sign Method (FGSM)~\cite{Goodfellow2015ExplainingAH}}
One popular method that has been used to generate adversarial examples in the wireless domain is the FGSM.
The FGSM is a one-step gradient-based method developed on finding the scaled sign of the gradient of the cost function and aims at minimizing the strength of the perturbation. It is a computationally effective method for adversarial attack generation based on the postulation that the DNNs are prone to linear-type adversarial attacks because of their use of linear techniques for easier optimization. The FGSM-generated perturbation is achieved by linearizing the model's cost function $J(w,x,y)$ around the current value of $x$. FGSM generates an adversarial example that significantly resembles the original input $x$ by starting with an original input to the ML model and adjusts it in the direction of the gradient of the loss function with respect to the input by an amount limited by a parameter $\epsilon$ stated as:
\begin{equation}
x^* = x + \epsilon \cdot \text{sign}\textcolor{black}{(}\nabla_x J(x, y,w)),
\end{equation}
where $x^*$ is the crafted adversarial example, $x$ is the original input, $J(x,y,w)$ is the cost function of the DNN with parameters $w$, $y$ is the class associated with $x$, and $\epsilon$ is a small number to limit the perturbation~\cite{fili2019AdExinRF}. Note that the Fast Gradient Method (FGM) is a generalization of FGSM to meet the $L_2~$norm bound ${|| x^* - x ||}_2 < \epsilon$ as
\begin{equation}
x^* = x + \epsilon \cdot \frac{\nabla_x J(x, y, w)}{{{|| \nabla}_x J(x, y, w)||}_2}.
\end{equation}

FGSM is also used in iterative methods such as the following.
\begin{itemize}
	\item \textit{Basic iterative method (BIM)~\cite{Kurakin2017AdversarialEI}:} The BIM can be viewed as multiple steps of FGSM with a small step size as well as a clipping of results after each iteration to establish that the results are in an $\epsilon$-neighborhood of the original input data. The iteration for the BIM is given by 
\begin{equation}
\label{eq:bim}
x_i^* =  \text{clip}_{x,\epsilon} \left(x_{i-1}^* + \epsilon \cdot \text{sign}(\nabla_x J({x^*_{i - 1}}, y, w))\right),
\end{equation}
where $\text{clip}_{x,\epsilon}$ clips the values of the adversarial sample to an $\epsilon$-neighborhood of the original sample $x$.
	\item \textit{Projected Gradient Descent  (PGD) \cite{Madry2018TowardsDL}:} PGD is a multi-step variant of FGSM on the negative loss function that helps initiate adversarial attacks with the aim of examining or analyzing the performance of a neural network from an optimization point of view. It puts a limit on the total perturbation in $L_{\infty}$-norm from $x^*_0 = x+ U_n(-\epsilon,\epsilon)$, where $U_n(-\epsilon,\epsilon)$ is the uniform random perturbation. In $t$ iterations, $x$ is determined as
\begin{equation}
\label{eq:pgd}
x^*_t = {\Pi}_{B_{\epsilon}(x)}  \left(x^*_{t-1} + \epsilon \cdot \text{sign}\left(\nabla_x J(x^*_{t - 1}, y, w)\right)\right) ,
\end{equation}
where ${\Pi}_{B_{\epsilon}(x)}$ is the projection operator. The procedure of adversarial example generation includes adding noise, computing gradient, stepping, and projecting back.

	\item \textit{Momentum Iterative Method (MIM)~\cite{Dong2018BoostAdAttM}:} The MIM~ was developed to solve the problems related to over-fitting and local minima. It helps speed up the gradient descent iterations. In scenarios where the performance of FGSM is inhibited due to the presence of strong interference and distortion, the MIM applies momentum to the iterative attacks so as to stabilize and maintain the correct direction of the update with generalization of adversarial examples.
\end{itemize}

\subsubsection{L-BFGS Method~\cite{Szegedy2014IntriguingPO}} Box-constrained L-BFGS is an example of optimization-based method for generating adversarial examples. For an instance of $x$, L-BFGS finds a different  $x^{*}$ that is very much alike to $x$ based on the $L_2$ distance but is labeled differently by the classifier. This problem is laborious and is modeled as the following constrained minimization:
\begin{eqnarray}
\label{eq:lbfgs}
& \min &{|| x - x^{*} ||}_2^2, \\ \nonumber
& s.t. & f_w(x+r_x) = y', \\ \nonumber
&& (x+r_x) \in [0,1]^n ,
\end{eqnarray}
where $y'$ is the target class. To solve (\ref{eq:lbfgs}), it is transformed to  
\begin{eqnarray}
\label{eq:lbfgssolved}
& \min & \lambda \cdot {|| x - x^{*} ||}_2^2 + J(x, y', w)\\ \nonumber
& s.t. & x^* \in [0,1]^n ,
\end{eqnarray}
where  $J(x, y^\prime, w)$ is a function mapping  an instance of $x$ to a positive real number using the loss function. In (\ref{eq:lbfgssolved}), the cross-entropy loss function is typically used as the loss function $J$ and line search is used to find the positive constant $\lambda$ that generates an adversarial example of minimum distance.

\subsection{\textcolor{black}{Generation of Exploratory Attack}}

~\textcolor{black}{In an exploratory attack, an adversary seeks to understand the internal operations on an ML model and leverage on such knowledge to attack the ML model and ultimately undermine the effectiveness of the ML-based system. This is achieved by developing a surrogate (or shadow) model that can mimic the operation of the original ML model. The approach is to steal training data and build models that have the same (or similar) output as the original model. With the information on how the original model functions, the attacker can make informed decisions as to how to attack and alter the overall performance of the ML-based system.} For example, in~\cite{Erpek2019DLforLM}, an exploratory attack is launched on the cognitive radio as a preliminary step to learn when and how to jam the cognitive radio transmissions. The adversary senses a channel, captures the transmitter’s decisions and learns the pattern of successful transmissions by tracking acknowledgments from the receiver, and launches an exploratory attack to build a DL classifier (surrogate model) that is functionally equivalent to the one at that transmitter. In another example~\cite{Sagduyu2019IoTNetSec}, an adversary observes the spectrum and develops a DNN to infer the channel access algorithm used by an IoT transmitter and predict the outcome of the transmissions before jamming it. Exploratory attacks are also launched to infer reinforcement learning mechanisms used in wireless channel access and jam the underlying communications.~\textcolor{black}{~\cite{Zhong2020adversarial, Wang2020adversarial} designed a DRL-based jamming attacker that employs a dynamic policy and aims to minimize the channel access accuracy of a DRL-based dynamic channel access user. The DRL attacker is able to observe the victim's interaction with its environment for a period of time and learns the activity pattern. The attacker has no prior information about the channel switching pattern on a victim's action policy and both DRL-based systems can interact with each other, retrain their models and adapt to the opponent's policy. Also,~\cite{Shi2021HowTA, Shi2021Adversarialaa} introduced a reinforcement learning (RL) algorithm to disrupt 5G network slicing  by observing the spectrum and developing an RL-based surrogate model. This model decides which resource blocks to jam in order to attain a high number of failed network slicing requests. Jamming a resource block causes a reduction in the RL algorithm's reward thus, affecting the model's performance as the reward is used to update the RL algorithm.}

\textcolor{black}{Although the exploratory attacks described above follow the same paradigm, there are no standard attack generation methods as we have in evasion attacks. All the methods use a common approach based on learning the internal workings of the original model, building a surrogate model and attacking the ML-based model using the knowledge from the surrogate model. Table~\ref{table:modelum} shows examples of previous works on exploratory attacks.}

\subsection{\textcolor{black}{Generation of Causative Attack}}

\textcolor{black}{The causative attack seeks to downgrade the ability of a ML system to perform optimally by injecting vulnerabilities into the training process. These injected vulnerabilities can be as a result of the adversary by intentionally meddling with the label of the data or by crafting and embedding unique learnable features. The model learns these unique features which causes it to deviate from its original objective, thus, causing considerable degradation in the ability of the model to make good predictions. The injection of vulnerabilities might be during the initial model development or the model re-training process. Since the attack occurs before model training, the contamination affects any type of ML model and efforts on parameter tuning yields little or no improvement on the model's ability to make good predictions~\cite{He2020ToSecTh}. As we have observed in the exploratory attack, there are no standard attack generation methods but the paradigm of poisoning the training data is consistent in all forms of causative attacks. Table~\ref{table:modelum} shows some previous work on causative attack.}

\section{Review of Adversarial Machine Learning in Wireless Communications}
\label{sec:amlreview}

Grouped according to the type of DL problem, we discuss and review various studies on AML in wireless communication and present them in Tables~\ref{table:modclass} (for modulation classification), Table~\ref{table:specsensing} (for spectrum sensing), Table~\ref{table:signalclass} (for signal classification), and Table~\ref{table:otherareas} (for other areas in the wireless communication domain).

\subsection{\textcolor{black}{Modulation Classification and Signal Classification:}}
\textcolor{black}{Modulation classification -- the process between the detection and demodulation of a signal -- is an important step towards developing an intelligent radio receiver and has been extensively studied~\cite{DobreSurvey2007}. Recently, the research in this area has gained new attention with the use of DL.~\cite{TosheaConvRMRNet} showed that DL algorithms for modulation classification using RF data outperform the use of expert features and higher order moments. AML for modulation recognition has been extensively using evasion and/or trojan attacks. ~\cite{Sadeghi2019AdAttDLRSC, Lin2020ThreatAdAtt, Ke2019AppAdExComm} showed the susceptibility of DL models to adversarial attacks on modulation classification under different scenarios such as data with noise and variation in SNR level. ~\cite{Sadeghi2019AdAttDLRSC, Flowers2020EvaAdEx} pointed out that AdExs for modulation classification are more effective than perturbations crafted by adding Gaussian noise. Table~\ref{table:signalclass} shows AML studies focused on modulation classification. Similar to the modulation classification problem, AML has been also applied to wireless signal classification. Signal classification aims to identify the transmitting signal in a spectrum. Table~\ref{table:modclass} highlights relevant work on AML for signal classification.}

\begin{table*}[htbp] \scriptsize
\caption{Review of previous works on AML attacks in modulation classification.}
\centering
\begin{tabular}{|m{1cm}|m{2cm}|m{6cm}|m{2cm}|m{1.5cm}|m{2.5cm}|m{5cm}|}\hline

\textbf{Paper} & \textbf{Attack Type} & \textbf{Objective} & \textbf{Attack Method} & \textbf{DL Problem} & \textbf{Data used}\\ \hline\hline

\cite{kim2020OTAAdAtt} & White-box, Black-box, Targeted, Untargeted, Evasion& Explore how to launch a realistic evasion attack by taking into consideration the channel effect as well as the power constraint at the adversary. & FGM, UAP & Modulation classification  & RML2016.10A \cite{o2016radio}  \\ \hline

\cite{Ke2019AppAdExComm} & Evasion & Test the attack efficiency of two types of adversarial attacks. & FGSM, Box-constrained L-BFGS & Modulation classification & RML2016.04C \cite{Oshea2016ModRec}  \\ \hline

\cite{Lin2020ThreatAdAtt}  & Evasion, Black-box, White-box &  Explore the  feasibility and effectiveness of adversarial attacks and determine the optimal perturbation level for attack invisibility and success. & FGSM, PGD, BIM, MIM & Modulation classification & RML2016.10A \cite{o2016radio}  \\ \hline

\cite{Davaslioglu2019TroAttWSC} & Trojan attack & Introduce a new attack that slightly manipulates training data by inserting Trojans (i.e., triggers) to a few training data samples and then activate them later in test time.  & Embed Trojans in training data and trigger it in test time & Modulation classification & RML2016.10A \cite{o2016radio} \\ \hline

\cite{Flowers2020EvaAdEx} & Evasion & Evaluate the vulnerabilities of raw I/Q based modulation classification. & FGSM & Modulation classification & RML2016.10A \cite{o2016radio}, three synthetic datasets using GNU radio  \\ \hline

\cite{HameedGlobalSIP, Hameed2019TheBD, Hameed2020newQM} & Evasion & Minimize the accuracy of the intruder in determining the modulation scheme used by a transmitter by perturbing the channel input symbols. & PGD & Modulation classification & Locally generated data \\ \hline

\cite{filipovic2019MitAdEx} & Gray-box & Mitigate adversarial examples in RF classifiers using autoencoder preprocessing. & FGSM  & Modulation classification & RML2018A \cite{Oshea2018OTADL} \\ \hline

\cite{Kim2020AdversarialAW} & Evasion White-box Targeted & Investigate the use of multiple antennas to generate multiple concurrent perturbations over different channel effects (subject to total power budget) to the input of DNN-based modulation classifier at a wireless receiver. & FGM & Modulation classification   & RML2016.10A \cite{o2016radio} \\ \hline

\cite{Sadeghi2019AdAttDLRSC} & Evasion, White-box, Black-box & Show the susceptibility of DL models to adversarial attacks and present practical methods for crafting adversarial examples in modulation classification. & PCA-based UAP, FGM-modified bisection search & Modulation classification & RML2016.10A \cite{o2016radio} \\ \hline

\cite{kim2020HowToMake} & Evasion, Targeted, White-box & Fool a DL-based eavesdropper by a cooperative jammer transmitting adversarial perturbation so as to hide 5G communications from the eavesdropper. & FGM & Modulation classification and 5G system & Locally generated data using MATLAB 5G toolbox \\ \hline

\cite{Del2020effects} & Evasion  & Extend the use of communication-aware evasion attack through the use of forward error correction (FEC). & Perturbation created by Adversarial Mutation Network & Modulation classification & Synthetic data generated using Liquid DSP  \\ \hline

\cite{Bair2019OnLimofTar} & Evasion, Untargeted, White-box, Targeted  & Examine how the DL-based automatic modulation classifier breaks down in the presence of an adversary with direct access to its inputs. & FGSM, MI-FGSM  & Modulation classification  & RML2016.10A \cite{o2016radio}. \\ \hline

\cite{KimChannelAware} &Evasion, Black-box, Untargeted, White-box, Targeted & Explore how to design realistic adversarial attacks in the presence of realistic channel effects and multiple classifiers at different receivers. & FGM, UAP & Modulation classification  & RML2016.10A \cite{o2016radio} \\ \hline

\cite{restuccia2020generalized} & White-box, Untargeted Targeted  & Provide a generalized wireless AML problem (GWAP) and experimental evaluation of AML attacks launched against wireless DL systems & AML waveform jamming, AML waveform synthesis & Modulation classification, Radio Fingerprinting & RML2018.01A \cite{Oshea2018OTADL}, 1000-device dataset of WiFi and ADS-B transmissions \\ \hline

\cite{Del2020investigating} & Evasion  & Introduce spectral deception loss metric implemented during training to make the spectral shape more in-line with the original signal.  & Adversarial Mutation Network (AMN)  & Modulation classification  & Locally generated data \\ \hline

\cite{Hameed2019CommunicationWI} & Evasion  & Minimize the accuracy of an intruder while still ensuring the successful decoding of a signal by the intended receiver. & PGD, uniform random noise of $L_2$-norm to a block of $128$ I/Q symbols & Defense against modulation detection  & Locally generated data \\ \hline

\textcolor{black}{\cite{yi2021gradientbasedAdvDeMod}} & \textcolor{black}{Evasion} & \textcolor{black}{Study the effect of adversarial attacks on ML-based AMC model using different data-driven subsampling strategies} & \textcolor{black}{Carlini-Wagner} & \textcolor{black}{Modulation classification} & \textcolor{black}{RML2016.10B~\cite{rml_datasets}}\\ \hline

\textcolor{black}{\cite{bahramali2021RobAdvAttAgDNN}} & \textcolor{black}{white-box, black-box} & \textcolor{black}{Present an input-agnostic adversarial attack that is undetectable and robust to removal} & \textcolor{black}{UAP} & \textcolor{black}{Modulation classification} & \textcolor{black}{RML2016.10A \cite{o2016radio}}\\ \hline

\end{tabular}
\label{table:modclass}
\end{table*}


\subsection{\textcolor{black}{Spectrum Sensing:}}
\textcolor{black}{Most of the AML studies focused on spectrum sensing and dynamic spectrum access (DSA) use exploratory attacks in which the attacker learns the operational pattern of the DNN model and builds a surrogate model which gives the same or similar output  as the original classifier when given the same input. Spectrum sensing and DSA are key components of cognitive radio systems for efficient discovery and use of spectrum to achieve situational awareness~\cite{Shi2018SpecDaPoiAML}. Given the broadcast and over-the-air nature of wireless communications, there are various attacks that attempt to undermine the spectrum sensing results. This can include causative attack to change the sensed features as well as priority violation attacks to violate priorities in channel access by pretending to have higher priority and transmitting during the sensing phase~\cite{Sagduyu2019IoTNetSec}. AML attacks on spectrum sensing typically involve an adversary that develops a surrogate model and can correctly predict the outcome of the original DL model. Such an adversary jams the successful transmissions, thus reducing the success rate of the original model~\cite{Shi2018AdDL4CogRaSec, Erpek2019DLforLM, Zhong2020adversarial}. As shown in~\cite{Shi2018AdDL4CogRaSec, Erpek2019DLforLM, Sagduyu2019AdDLOTA, Luo2020attackers}, these attacks are more energy-efficient and more effective than random jamming. Table~\ref{table:specsensing} highlight the studies on AML for spectrum sensing and DSA applications.}


\begin{table*}[htbp] \scriptsize
\caption{Review of previous works on AML attacks in spectrum sensing.}
\centering
\begin{tabular}{|m{1cm}|m{2cm}|m{6cm}|m{2cm}|m{1.5cm}|m{2.5cm}|m{5cm}|}
\hline
\textbf{Paper} & \textbf{Attack Type} & \textbf{Objective} & \textbf{Attack Method} & \textbf{DL Problem} & \textbf{Data used}\\ 
\hline\hline

\cite{Shi2018AdDL4CogRaSec, Erpek2019DLforLM} & Exploratory & Launch an exploratory attack on cognitive radio as a preliminary step before jamming.  & Jam legitimate user transmissions & Cognitive radio channel access & Locally generated data \\ \hline

\cite{Shi2018SpecDaPoiAML} & Spectrum poisoning, Exploratory & Launch an over-the-air attack to infer the transmitter's behavior and falsify the spectrum sensing data by transmitting perturbations over the air. & Inject perturbation to the channel & Spectrum sensing & Locally generated data \\ \hline

\cite{Sagduyu2019IoTNetSec} & Exploratory, Evasion, Causative, Black box & Build a DNN classifier to infer the channel access algorithm of a transmitter and predict the transmission result, and develop a defense mechanism to confuse the adversary in its attack strategy. & Jamming, Spectrum poisoning, Priority violation & Spectrum sensing & Locally generated data \\ \hline

\cite{Luo2020attackers} & Black-box, Exploratory & Propose Learning-Evaluation-Beating(LEB) attack against cooperative spectrum sensing (extension of spectrum sensing data falsification attack on cognitive radio). & LEB attack & Cooperative spectrum sensing & Dataset collected from 5282 locations~\cite{Saeed2017LocLow}, Data from 5 locations using USRP N210 \\ \hline

\cite{Wang2020adversarial} & Black box Gray box & Develop deep reinforcement learning (DRL)-based dynamic channel access system and DRL-based jamming attack to study the sensitivity of DRL-based wireless communications to adversarial attacks. & DRL-based attacker use dynamic policy to minimize the victim's channel access & Dynamic spectrum access (DSA) & Locally generated data. \\ \hline

 \cite{Sagduyu2019AdDLOTA} & Exploratory, Evasion, Causative  & Launch over-the-air spectrum poisoning attacks by learning first the transmitter's behavior. & Attacker jams the transmitter's sensing period & Spectrum sensing, DSA & Locally generated data \\ \hline

\cite{Zhong2020adversarial} & Exploratory  & Compare the performances of two adversarial policies: feed-forward neural network (FNN) and DRL policies on the performance of a DRL-based dynamic channel access agent. & Jamming of a victim model & Dynamic spectrum access  & Locally generated data  \\ \hline

\textcolor{black}{\cite{sagduyu2021AMLfor5GCommSoc}} & \textcolor{black}{Exploratory, Evasion}  & \textcolor{black}{Present AML attack on spectrum sharing of 5G communications with incumbent users.} & \textcolor{black}{Manipulate inputs to the classifier by transmitting during spectrum sensing} & \textcolor{black}{Spectrum sharing} & \textcolor{black}{Locally generated data}\\ \hline

\end{tabular}
\label{table:specsensing}
\end{table*}

\subsection{\textcolor{black}{Other Areas in Wireless Communication:}}

\textcolor{black}{AML attacks have been also applied to other areas in wireless communications. This includes resource allocation such as network slicing~\cite{Shi2021HowTA, Shi2021Adversarialaa}, IoT data fusion process~\cite{ZluoPartialAttack}, waveform/communications as in power control \cite{manoj2021AdvAttDLBasPowAll}, beam alignment and initial access~\cite{kim2021adversarial}. A review of AML studies on other areas in wireless communication is presented in Table~\ref{table:otherareas}.}


\begin{table*}[htbp] \scriptsize
\caption{Review of previous works on AML attacks on signal classification.}
\centering
\begin{tabular}{|m{1cm}|m{2cm}|m{6cm}|m{2cm}|m{1.5cm}|m{2.5cm}|m{5cm}|}

\hline

\textbf{Paper} & \textbf{Attack Type} & \textbf{Objective} & \textbf{Attack Method} & \textbf{DL Problem} & \textbf{Data used}\\ 
\hline\hline
\cite{Shi2020OTAMIA} & Black-box, Exploratory & Show that DL-based signal classifiers are vulnerable to privacy threats due to over-the-air information leakage of their model. & Membership inference attack (MIA) & Wireless signal classification  & Locally generated data \\ \hline

\cite{Kim2020surrogatemodel} & Exploratory, White-box, Targeted & Investigate the channel effects on surrogate model built by an adversary that uses the over-the-air observed spectrum data.  & Maximum Received Perturbation Power (MRPP) attack \cite{kim2020OTAAdAtt} & Wireless signal classification & RML2016.10A \cite{o2016radio} \& locally generated data \\ \hline

\cite{fili2019AdExinRF} & Evasion & Develop methods for detection of adversarial perturbations used against wireless signal classifiers. & FGSM & Survey application \& spectrum sensing & RML2018A \cite{Oshea2018OTADL}, over-the-air Bluetooth, WiFi, ZigBee datasets\\ \hline

\cite{karunaratne2020penetrating} & Evasion, Untargeted, White-box & Fool a DL-based signal classifier (authenticator). & Reinforcement learning-based attack & Signal authentication & Simulated data and testbed data from Pluto SDRs  \\ \hline 

\cite{fili2019TargetAdEx} & Targeted, Black box & Analyze the effect of adversarial examples against Deep RF classifiers and also develop several defense mechanisms.  & Carlini-Wagner attack  & Signal \& , protocol classification & Locally generated WiFi, Bluetooth \& ZigBee data. RML2018A \cite{Oshea2018OTADL}\\ \hline

\textcolor{black}{\cite{bahramali2021RobAdvAttAgDNN}} & \textcolor{black}{white-box, black-box} & \textcolor{black}{Present an input-agnostic adversarial attack that is undetectable and robust to removal.} & \textcolor{black}{UAP} & \textcolor{black}{Signal Detection in OFDM Systems} & \textcolor{black}{RML2016.10A \cite{o2016radio}}\\ \hline

\end{tabular}
\label{table:signalclass}
\end{table*}


\begin{table*}[htbp] \scriptsize
\caption{Review of other previous works on AML attacks in wireless communications}
\centering
\begin{tabular}{|m{1cm}|m{2cm}|m{6cm}|m{2cm}|m{1.5cm}|m{2.5cm}|m{5cm}|}
\hline
\textbf{Paper} & \textbf{Attack Type} & \textbf{Objective} & \textbf{Attack Method} & \textbf{DL Problem} & \textbf{Data used}\\ 
\hline\hline

\cite{Sadeghi2019PhyAdAttAuto} & Black-box, Exploratory, Evasion & Show that end-to-end learning of communication systems through autoencoder can be extremely vulnerable to physical adversarial attacks. & Iterative method based on UAP & End-to-end autoencoder communication systems & Locally generated data \\ \hline

\cite{ZluoPartialAttack} & Exploratory  &  Employ AML techniques to launch attacks against the IoT data fusion process. & Partial-model attack & IoT data fusion/aggregation process  & Synthetic data from Gaussian distributions \\ \hline

\cite{Liu2020AdvAttonDL} & White-box, Untargeted & Present a method to craft adversarial attack and show its effect on DL-based channel state information (CSI) feedback process.  &  Attacker modeled by a bias layer between the encoder and decoder of CsiNet & Massive MIMO CSI feedback process  & Locally generated data generated using COST 2100 channel model \cite{LiuCost2100model} \\ \hline

\cite{Wang2020defense} & Black-box  & Develop defense strategies against DRL-based jamming attacks on a DRL-based dynamic multichannel access agent.  & DRL-based jamming attacker  & Multichannel access  & Locally generated data \\ \hline

\textcolor{black}{\cite{Shi2021HowTA}} & \textcolor{black}{Exploratory}  & \textcolor{black}{Present an over-the-air attack to manipulate the RL algorithm and disrupt 5G network slicing.} & \textcolor{black}{Adversary builds an RL-based model and jams selected resource blocks} & \textcolor{black}{Resource allocation for 5G network slicing} & \textcolor{black}{Locally generated data} \\ \hline

\textcolor{black}{\cite{Shi2021Adversarialaa}} & \textcolor{black}{Exploratory}  &  \textcolor{black}{Present flooding attack on 5G network slicing.} & \textcolor{black}{Adversary crafts fake network slicing requests to consume the 5G radio access network resources} & \textcolor{black}{Resource allocation for 5G network slicing} & \textcolor{black}{Locally generated data}\\ \hline

\textcolor{black}{\cite{sagduyu2021AMLfor5GCommSoc}} & \textcolor{black}{Exploratory, Spoofing}  &  \textcolor{black}{Present adversarial attack on signal authentication in network slicing.} & \textcolor{black}{Spoofing attack on the network slicing application} & \textcolor{black}{Physical layer authentication of 5G device} & \textcolor{black}{Locally generated data}\\ \hline

\textcolor{black}{\cite{kim2021adversarial}} & \textcolor{black}{non-targeted}  &  \textcolor{black}{Investigate the vulnerability of a DNN used for mmWave beam prediction as part of the initial access process in 5G and beyond communications.} & \textcolor{black}{FGM, adversary tries to change the beam to one of the worst beams} & \textcolor{black}{mmWave beam prediction in 5G} & \textcolor{black}{Locally generated data}\\ \hline

\textcolor{black}{\cite{manoj2021AdvAttDLBasPowAll}} & \textcolor{black}{white-box, black-box}  &  \textcolor{black}{Show that adversarial attacks can break DL-based power allocation in the downlink of a massive multiple-input-multiple-output (maMIMO) network.} & \textcolor{black}{(FGSM), momentum iterative FGSM, and PGD} & \textcolor{black}{Power allocation in a maMIMO} & \textcolor{black}{DL power allocation in massive MIMO dataset~\cite{sanguinetti2019deep}}\\ \hline

\end{tabular}
\label{table:otherareas}
\end{table*}

\section{Detection and Mitigation of AML Attacks in Wireless Communications}
\label{sec:detectmiti}
Next, we present various methods that have been used to defend against AML attacks on wireless communication systems.

\subsection{Train with Adversarial Perturbations}
Training the DNNs with adversarial examples is a popular method of mitigating the effect of adversarial attacks and ensuring robustness~\cite{Madry2018TowardsDL}. Adversarial training consists of generating adversarial examples according to one or more attack methods and retraining the DNNs with labeled adversarial examples. The goal is to prevent an adversary from identifying the structure of the signals transmitted so as to prevent it from disrupting the communication system. This defense mechanism has been applied to protect wireless signal classifiers against evasion attacks~\cite{KimChannelAware}, where randomized smoothing \cite{Cohen2019certified} used in training time increases the robustness of classifiers against adversarial attacks later in test time. 

There are some challenges that pertain to adversarial training; some of these include a scenario where the adversary uses a different attack from the one used in training. In addition to that, an adversary can create new perturbations using a model trained with adversarial examples. However, training with adversarial examples typically reduces the performance of DL models on unperturbed signals. In trying to overcome these challenges, training can be performed with multiple adversarial examples~\cite{Madry2018TowardsDL}. In addition, certified defense~\cite{Raghunathan2018certified} can be built in train and test times to ensure statistical significance of classification results in the presence of adversarial attacks in test time. Certified defense has been used in~\cite{KimChannelAware} to provide performance guarantees for wireless signal classifiers in the presence of adversarial examples. Another way to mitigate adversarial attacks during training is pre-training of the DL based wireless signal classifiers by an autoencoder such that the trained model becomes robust to the deceiving effect of adversarial examples \cite{filipovic2019MitAdEx}.

\subsection{Statistical Approaches}
\begin{itemize}
	\item \textit{Peak-to-average power ratio (PAPR) of RF samples:} One way of detecting adversarial examples it to leverage the properties of radio frequency (RF) data and apply a statistical test by taking advantage of the PAPR of digitized RF samples of the received signals~\cite{fili2019AdExinRF}. The PAPR distribution is used in the wireless domain to depict a specific class of signal's footprint such that if the inferred output of an input signal is categorized to a particular class and its PAPR statistic contradicts with high level of confidence, further analysis and assessment should be carried out to reach a true conclusion. To determine if an adversarial example is present  in a wireless communication environment, the Kolmogorov-Smirnov (KS) two-sample test (performed by collecting a sample sized input and then computing and evaluating the PAPR distributions for the samples) can be used to ascertain if the PAPR is similar to the statistic of an adversarial example or a legitimate example~\cite{fili2019AdExinRF}. 
	
	\item \textit{Use the softmax outputs of the DNN classifier:} This is also a KS statistical test for adversarial example detection. The softmax output of the DNN classifier is used to determine if adversarial examples have brought about a change in distribution using the statistics of the final layer of a neural network. Typically, the statistical size of the input data is a good determination of the quality of the statistical test. The effectiveness of the method depends on the type of waveform and propagation channel, as discussed in~\cite{fili2019AdExinRF}.
	
	\item \textit{Statistical methods to detect adversarial triggers:} Adversarial triggers such as Trojans (or backdoors) inserted to training data can be detected by statistical methods such as clustering and median absolute deviation (MAD) algorithm. In particular, the MAD algorithm computes the median of the absolute deviations from the data's median, namely $\text{median}(|x_i - \hat X|)$, where $\hat X = \text{median}(X)$, to discover outliers, and is shown in \cite{Davaslioglu2019TroAttWSC} to be effective against Trojan attacks on wireless signal classifiers. 
\end{itemize}

\subsection{Randomizing the DL Algorithm by Adding Small Variations to Classifier Outputs}

This approach adds slight variations to the output of the DNN classifier and deliberately performs some (typically a small number of) erroneous actions so as to deceive the adversary and defend against exploratory attacks. For example, the transmitter can intentionally utilize a busy channel for transmitting or skip a transmission opportunity in an idle channel such that the adversary that observes the spectrum cannot train a reliable surrogate model. This defense mechanism has been applied to protect wireless signal classifiers against inference attacks (before launching jamming attacks) in \cite{Erpek2019DLforLM} as well as against evasion and causative attacks in \cite{Sagduyu2019AdDLOTA}. Note that network uncertainty (e.g., due to dynamic channel and traffic effects) and randomization or obfuscation-based defense methods have also been leveraged before against traditional network attacks such as jamming and traffic inference to confuse the adversaries \cite{xu2005feasibility, sagduyu2011jamming, sagduyu2010wireless, hou2020proto}. Defense against AML attacks on reinforcement learning follows a similar diversification approach to protect wireless communications \cite{Wang2020adversarial, Wang2020defense}. The defender may alternate among different channel access decisions (derived through  proportional-integral-derivative (PID) controller and imitation learning) to increase the uncertainty at the adversary. Another defense approach is to adopt orthogonal policies in reinforcement learning to prevent the adversary from quickly switching between its imitated policies. On top of these mitigation policies, it is also viable to detect the adversaries by monitoring changes in rewards and distinguishing the spectrum environment changes from adversarial jamming attacks \cite{Wang2020adversarial, Wang2020defense}. Table~\ref{table:modelum2} reviews the defense strategies that have been used by previous works in the wireless domain.

\begin{table} [ht] \scriptsize
\caption{Summary of defense strategies against AML attacks in wireless communications.}
\centering
\begin{tabular}{|m{1.5cm}|m{6.5cm}|}
\hline
\multicolumn{1}{|c|}{\textbf{Paper}}& \multicolumn{1}{|c|}{\textbf{Defense Strategy}} \\
\hline \hline
\multirow{1}{*} ~\cite{Sagduyu2019IoTNetSec, Shi2018AdDL4CogRaSec, Erpek2019DLforLM, Sagduyu2019AdDLOTA, Hameed2019CommunicationWI} & Deliberately makes a small number of incorrect transmit actions so as to fool the adversary and prevent it from building a reliable classifier, e.g., transmit in a busy channel. \\ \hline
\multirow{1}{*} ~\cite{Hameed2019TheBD, HameedGlobalSIP} & Slightly modifies the transmitted signal; enough to fool an intruder and allow for correction using the error correction modules of the receiver. \\ \hline
\multirow{1}{*} ~\cite{filipovic2019MitAdEx} & Uses an autoencoder for the DL classifier to remove non-salient features thereby removing perturbations. \\ \hline
\multirow{1}{*} ~\cite{Davaslioglu2019TroAttWSC} & Uses the MAD algorithm and two-step technique of dimensionality reduction and clustering via t-distributed stochastic neighbor embedding (t-SNE) to detect Trojan triggers.\\ \hline
\multirow{1}{*} ~\cite{fili2019AdExinRF, kokaljfilipovic2019adversarial, fili2019TargetAdEx} & Applies KS Statistics that uses the peak-to-average power ratio or the softmax output of ML model for distribution shift detection. \\ \hline
\multirow{1}{*} ~\cite{KimChannelAware} &  Applies randomized smoothing in training time to increase the robustness of classifiers against adversarial attacks in test time and ensures statistical significance of classification results in the presence of adversarial attacks in test time.\\ \hline
\multirow{1}{*} ~\cite{Wang2020adversarial, Wang2020defense} &  Applies a diversification approach by alternating the reinforcement learning algorithm among different channel access decisions (via PID controller and imitation learning) and adopting orthogonal policies to restrict the adversary's response.  \\ \hline
\multirow{1}{*} ~\cite{Hameed2019CommunicationWI} & Adding a constrained perturbation to the signal and ensuring that the signal remains decodable by the oblivious legitimate receiver and minimizing detection accuracy at the intruder. \\ \hline
\multirow{1}{*} ~\cite{fili2019TargetAdEx} & Pre-training the DL classifier using an autoencoder which is expected to filter out non-salient features that may slightly lower the accuracy of classification and makes the classifier more robust to adversarial or environmental corruption.\\ \hline
\multirow{1}{*} ~\cite{Luo2020attackers} & Introduced influence-limiting defense by using decision-flipping influence which indicates the probability of finding a malicious input that modifies the decision output given the unperturbed data by changing part of the unperturbed signal. \\ \hline
\end{tabular}
\label{table:modelum2}
\end{table}

\section{Future Outlook of Adversarial \textcolor{black}{Machine Learning} in Wireless Communications}
\label{sec:futureoutlook}

As ML/DL becomes the core of current and emerging communications systems such as 5G and 6G, ML/DL itself becomes susceptible to adversarial effects. To build on the promise of intelligent operations and efficient resource management with the application of ML in dynamic wireless environments, it becomes imperative to develop ML models that are secure, resilient and robust to attacks by taking into account the unique properties of wireless communications. An AML attack model that is carefully characterized by considering wireless properties is expected to be the foundation of ML driven wireless security research and development for the existing and emerging communication systems. In this section, we discuss some important ideas that can help in realizing the promise of robust models that are effective even in the presence of adversaries.\\

\begin{itemize}
	\item \textcolor{black}{\textit{Wireless communication dataset:} The limitation of standardized real-world datasets that adequately represent real life scenarios in the wireless communication domain is a challenge that needs to be solved. Compared to other domains such as CV and NLP, there are just a handful of publicly available ML/DL datasets in the wireless domain to work with such as~\cite{Oshea2016ModRec, OsheaOverTheAir2018, GenesysLab, al2020massive, Alkhateeb2019, Ezuma2020}. Typically, these datasets do not include adversarial effects recorded. It is important for the research community to develop more publicly available datasets, representing different scenarios including not only variations in channel, interference type and waveform, but also AML attacks. This approach will help with benchmarking solutions for robust development and evaluation of the DL models for wireless versions of AML attacks. From literature reviewed, a large ratio of the work on AML in wireless communication is in the area of modulation classification. This is largely because modulation classification datasets (e.g.~\cite{Oshea2016ModRec, OsheaOverTheAir2018}) are the seemingly standardized datasets for Ml/DL studies in the wireless communications domain. Although more recent datasets are available such as~\cite{ GenesysLab, al2020massive, Alkhateeb2019, Ezuma2020} as well as testbed efforts such as ~\cite{Breen+:wintech20} expand datasets, it is critical to implement the AML attacks and corresponding defenses with embedded radio platforms to assess the real channel and radio hardware effects.}

	\item \textcolor{black}{\textit{Robust features:} The need for robust features in developing ML/DL models in wireless communication is also an important consideration for future systems. ~\cite{ilyas2019adversarial} laid the analytical foundation and showed via experimental data that adversarial attacks are very effective because most existing ML/DL models are not developed using robust features. Although the discussion of feature engineering and identifying robust features are on-going especially in CV and NLP domains, new techniques must be developed to identify the most important features which are robust to the effect of adversarial attacks in wireless communication systems. An important research direction for AML in wireless communications is to identify robust features in datasets, disentangle non-robust features from robust features, train models that generalize well, and are robust to adversarial attacks in wireless communications.}

	\item \textcolor{black}{\textit{Certifiable defense:} The design and development of certifiable defense mechanisms for AML is another important component of the work that needs to be achieved to realize the full potential of ML/DL in wireless communications domain. Various defense mechanisms have been proposed, including adversarial training and use of statistical methods but most of these defenses become ineffective with more powerful adversaries. This has led to the need for certifiable defense mechanisms, where prediction on the test data can be verified to be a constant within a small region around the input data. Although there are some efforts that present certifiable defense using randomized smoothing~\cite{KimChannelAware} with reasonable performance, more rigorous evaluation of such approaches as well as development of other methods are needed to provide security guarantees for ML models used in wireless communications.}
\end{itemize}

	\textcolor{black}{Although important areas of development to ensure good performance from robust ML/DL models have been discussed in this section, for the safe adoption of ML/DL-based models into future communication systems, there is also a need to move from black-box models to interpretable models. Such paradigm as seen today includes physics-guided ML~\cite{RaiPhysicsGuidedMLIEEEAccess2020}, where the fundamental laws of physics is infused into the learning process to improve generalizability and explainability, and ensure robustness to attacks, as needed in wireless communication systems.}

\section{Conclusion}
\label{sec:conclusions}
\textcolor{black}{The recent trends observed in this survey show that the AML attacks can effectively disrupt wireless communications. Compared to the traditional jamming attacks or the addition of noise/interference, adversarial attacks are more difficult to detect operating with small spectrum footprint, more energy efficient, and overall more effective. Furthermore, as wireless communications evolves into smarter systems as in 5G and 6G, the AML attacks also become smarter. The attacks have evolved from adversaries that send jamming signals randomly to overwhelm a communication system to those that develop surrogate models to know the internal workings of the original model and send out small perturbation signals to fool the ML/DL models and disrupt wireless communications.}

The goal of AML is to support safe adoption of ML/DL solutions to the emerging applications in the presence of adversaries. Wireless communications strongly benefit from ML/DL applications in terms of learning from spectrum data and solving complex tasks. Various studies reviewed in this paper have established that AML attacks on the ML/DL-driven wireless systems are effective and pose serious threats leading to major performance degradation. This review attempts to present the current state of the art in the area of AML attacks specific to the wireless communications domain and help the research community identify the research accomplishments as we forge ahead to investigate methods by which efficient data driven intelligent systems can be built while taking into account the unique nature of wireless communications, robustness, resilience, and security of ML/DL based methods. We summarized the latest research efforts by reviewing the attack types, adversarial example generation, and defense strategies, established a taxonomy across different attack categories, and discussed the future outlook to AML in wireless communications. This is an emerging research area and there is an urgent need for further investigations to quantify the impact of AML attacks and detect and mitigate them to enable the safe adoption of the promising ML/DL solutions for wireless communications.


\bibliographystyle{IEEEtran}
\bibliography{AdversarialLearn_updated}
\end{document}